# Observation of Quantum Spin Hall States in InAs/GaSb Bilayers under Broken Time-Reversal Symmetry

Lingjie Du[1], Ivan Knez[1,2], Gerard Sullivan[3], and Rui - Rui Du[1]

[1]Department of Physics and Astronomy, Rice University, Houston, Texas 77251-1892, USA

[2]IBM Research – Almaden, San Jose, California 95120, USA

[3]Teledyne Scientific and Imaging, Thousand Oaks, CA 91630, USA

**Topological insulators (TIs) are a novel class of materials with nontrivial surface or edge states[1,2]. Time-reversal symmetry (TRS) protected TIs are characterized by the $Z_2$ topological invariant[3,4,5,6] and their helical property becomes lost in an applied magnetic field. Currently there exist extensive efforts searching for TIs that are protected by symmetries other than TRS. Here we show, a topological phase characterized by a spin Chern topological invariant[7,8,9,10] is realized in an inverted electron-hole bilayer engineered from indium arsenide-gallium antimonide (InAs/GaSb) semiconductors which retains robust helical edges under a strong magnetic field. Wide conductance plateaus of $2e^2/h$ value are observed; they persist to 12T applied in-plane magnetic field without evidence for transition to a trivial insulator. In a perpendicular magnetic field up to 8T, there exists no signature to the bulk gap closing. While the Fermi energy remains inside the bulk gap, the longitudinal conductance increases from $2e^2/h$ in strong magnetic fields suggesting a trend towards chiral edge transport. Our findings are first evidences for a quantum spin Hall (QSH) insulator protected by a spin Chern invariant. These results demonstrate that InAs/GaSb bilayers are a novel system for engineering the robust helical edge channels much needed for spintronics and for creating and manipulating Majorana particles in solid state.**

Topological order is a new paradigm in classification of condensed matter systems, describing certain system observables, such as charge or spin conductance, via topological invariants, *i.e.* distinct system characteristics which remain unchanged under smooth deformations of its band structure. Besides topological considerations, time reversal symmetry has been widely believed to be a necessary ingredient for emergence of the QSH insulating phase, commonly characterized via the $Z_2$ topological invariant. More recently, it has been suggested that robust QSH states may arise in systems without TRS. In this case, the system is characterized via spin Chern invariants (SCI), which remain quantized under broader symmetry conditions[7,8,9], including inversion, particle-hole symmetry or even in the absence of any underlying symmetries apart from charge conservation[10] – resting solely on the topology of the band structure and smoothness of the confining potential.

Specifically, for the QSH state spin Chern numbers equal $C_{\pm}=\pm 1$ for the spin up and spin down sectors respectively, giving total spin Chern number $C_S = C_+ - C_- = 2$, and quantized spin Hall conductance of $C_S \cdot e/4\pi$ for systems with $S_z$ spin conservation. Similarly, charge transport is described via total charge Chern number $C_Q = C_+ + C_- = 0$ and respective Hall conductance equal to $C_Q \cdot e^2/2h = 0$. As a result, the QSH state manifests via counter-propagating edge channels giving quantized spin Hall conductance and net zero Hall conductance. Under a perpendicular magnetic field, edge channels of one spin component annihilate as $C_\pm$ changes to either +1, 0, or 0, -1, giving $C_S = 1$ and $C_Q = +1$ or -1 with Hall conductance now quantized to $\pm e^2/h$ as illustrated in Figure 1a. The described transition from QSH state to quantized Hall state, and accompanied change in $C_\pm$, occurs only upon closing and re-opening of either the bulk energy gap or the spectrum gap of the projected $S_z$ spin operator[9,10].

The spin Chern $C_S = 2$ insulating state is here realized in InAs/GaSb quantum wells where electron-hole bilayer naturally occurs due to the unique broken-gap band alignment of InAs and GaSb[11]. In particular, the conduction band of InAs is some 150 meV lower than the valence band of GaSb, which results in charge transfer between the two layers, and emergence of coexisting 2D sheets of electrons and holes, trapped by wide gap AlSb barriers. The positions of the electron and hole subbands can be altered by changing the thickness of InAs and GaSb layers, resulting in topologically trivial and non-trivial energy spectra shown in Figure 1d for narrower wells and wider wells, respectively[12,13]. In addition, due to the charge transfer and resulting band bending, both the topology of the band structure as well as the position of the Fermi energy can be continuously tuned via front and back gates[12] as shown in Figure 1b.

In the topologically non-trivial regime electron-hole subbands cross for some wave-vector values $k_{cross}$ and due to the tunneling between the wells, electron and hole states hybridize, lifting the degeneracy at $k_{cross}$ and opening an inverted energy gap[13,14] on the order of 40 - 60K. It has been proposed[15] that at the edge of the system, this inverted mini-gap must close before the normal gap such as in nearby native oxides or vacuum can open, necessitating a linearly dispersing helical edge spectrum. Some evidence for such helical states has been previously reported[16,17], albeit their unequivocal identification has been limited due to finite bulk density of states in the mini-gap, resulting from disorder broadening and imperfect hybridization of electron-hole levels[18,19].

Remarkably, as shown here, these 2D bulk states can be localized[20] even at finite temperatures by Si dopants of a relatively small density (~$1 \cdot 10^{11}$cm$^{-2}$, equivalent to 1,000 atoms in 1μm x 1μm device) at the interface, which serve as donors in InAs and acceptors in GaSb, creating a mobility gap of ~26K in the bulk energy spectrum. On the other hand, because the edge states are topological in nature, the disorder has very little effect on their existence and transport properties. In fact, as the Fermi energy is tuned into the mobility gap via front gates, four-terminal longitudinal conductance measurements in $\pi$-bar configuration, as well as six-terminal measurements in the Hall bar geometry for the mesoscopic 2μm x 1μm samples, reveals wide plateaus near perfectly quantized to $4e^2/h$ and $2e^2/h$, respectively (Figure 2a), as expected for helical edge channels[6,16] based on Landauer-Buttiker analysis[21]. Note that the conductance value here is quantized to better than one percent - unprecedented by any other known topologically ordered system other than Integer and Fractional Quantum Hall Effects (IQHE and FQHE), indicating a high degree of topological protection. In fact, besides the classical Chern number which characterizes IQHE, the spin Chern number is the only other topological invariant which has been rigorously proved to be robust to disorder[9], indicating that the InAs/GaSb bilayer system manifestly belongs to spin Chern insulating phase.

Furthermore, as the length of the Hall bar is increased to macroscopic dimensions, longitudinal resistance in the mobility gap linearly increases with the device length. It has been previously shown [6, 16, and ref. there-in] that helical edge states are protected only against elastic back-scattering, while inelastic scattering gives an upper critical length $\lambda_\varphi$ at which edge transport breaks down and counter propagating spin-up and spin-down channels equilibrate. In this case, approximate longitudinal resistance is obtained by series addition of $N \sim L/\lambda_\varphi$ half-quantum resistors[21], giving a total resistance value of $(L/\lambda_\varphi) \cdot h/2e^2$. This approximation is in excellent agreement with the data presented in Figure 2d, giving $\lambda_\varphi = 4.4$ μm. We emphasize that finite resistance values observed in the mobility gap in Figure 2, stem only due to the edge states, which have extended character, while the bulk states at the Fermi level are completely localized.

We note that in the context of IQH, a precisely quantized Hall plateau (in quantum resistance $h/e^2$) is due the opening of a localization gap in the Landau level spectrum[22,23]; here the existence of wide conductance plateau should be attributed to the opening of a mobility gap from Si-doping. This is confirmed via capacitance measurements as well as transport measurements in the Corbino disk geometry presented in Figure 3. Total capacitance of the system is most simply modeled as a series combination of geometric and quantum capacitances, where the latter is directly proportional to the density of states [13 and ref. there-in]. As a result, measuring the total capacitance between the front gate and 2D electron-hole bilayer, provides an indirect measure of the density of states of the system. Figure 3a shows *C-V* plots for sample temperatures varied from 45K down to 300 mK. As the temperature is reduced below ~30K, the *C-V* characteristic develops a wide dip corresponding to reduced DOS within the mini-gap. An additional dip in capacitance is observed as the temperature is further reduced below ~10K, indicating opening of the mobility gap in the energy spectrum; a wide conductance plateau emerges only in this regime.

The energy scale of the 2D mobility gap is quantitatively determined from transport measurements in Corbino samples, shown in Figure 3, as a function of temperature and magnetic field. In Corbino disk geometry, edge transport is shunted via concentric contacts, and hence conductance measurements probe bulk properties exclusively. In this case, transverse conductance is suppressed to zero in the mobility gap, showing exponentially activated temperature dependence and allowing direct extraction of mobility gap values (see Supplementary Information). As shown in Figure 3e, the mobility gap increases from ~26K at zero magnetic field to ~40K at 6T. In consequence, at temperatures on the order of 1K and below, the system is completely bulk insulating and transport occurs only along the edge. As a result, quantized conductance in mesoscopic structures and finite resistance values in longer devices shown in Figure 1 are solely a property of the topological edge channels.

Finally we study the edge transport properties under TRS breaking by applying magnetic fields along each major axis of the device. Unexpectedly, under in-plane magnetic fields applied along and perpendicular to the current flow, the mobility-gap conductance plateau value remains quantized for both two-terminal and four-terminal mesoscopic samples, as well as stays constant for longer devices, for fields even up to 12T (Figure 4a and 4b). This unequivocally demonstrates that edge transport in InAs/GaSb bilayers is dramatically robust to TRS breaking, positing strong topological protection by SCI. On the other hand, for perpendicular fields, Figure 4c and 4d, the longitudinal conductance plateau value gently increases in four-terminal devices and similarly decreases in two-terminal sample indicating a trend to chiral transport regime, as illustrated in Figure 1a. Note that in the chiral transport limit[22], there is no longitudinal voltage drop in four-terminal geometry, while in the two-terminal case conductance is quantized to $e^2/h$. While transition from helical to chiral transport is generally accompanied by reduction in spin Chern number and hence gap closing and re-opening, we note that the bulk energy mobility gap remains open as shown in Figure 3d, suggesting a still higher magnetic field would be needed to reach a topological transition. Possible opening and closing of the spectrum gap[9,10] of the projected $S_z$ spin operator in InAs/GaSb bilayers may be separately studied in the future.

In conclusion, in the mobility gap of InAs/GaSb bilayer system, we observe Quantum Spin Hall state quantized to a remarkable degree of accuracy and highly robust to TRS breaking, indicating strong topological protection via SCI. Realization of a spin Chern insulating phase in this system uncovers a new family of topological phases and paves way towards more robust spintronics and quantum information devices.

**Figure Captions**

**Figure 1 a.** Topological phases and topological phase transitions are protected by spin Chern number, as marked below the panels corresponding to quantum spin Hall (left) and quantum Hall (right) states, where helical or chiral edges are shown. There is a wide range of magnetic fields between these two topological phases, where the experimental results are reported here. **b.** depicts the device structures, **c.** shows schematically the band structure of the InAs/GaSb bilayers, and the potential fluctuations induced by Si dopants at the interface. **d.** The helical edges in an inverted bilayer where the edge states must cross to form a 1D Dirac dispersion.

**Figure 2 a.** Wide conductance plateaus quantized to $2e^2/h$ and $4e^2/h$, respectively for two device configurations shown in inset; both have length 2 um. **b.** Plateau persists to 4K, and conductance increase at higher temperature due to delocalized 2D bulk carriers. **c.** Electrical charge transport in large devices is due to edge channels, and **d.** the resistance scales linearly with the edge length, indicating a phase coherence length of 4.4 um; the coherence length is independent of temperature between 0.02K and 4K.

**Figure 3 a.** Capacitance-gate voltage curves are shown. As the temperature is reduced below ~30K, *C-V* characteristic develops a wide dip corresponding to reduced DOS within the electron-hole-hybridization induced mini-gap. around the center of the mini-gap, additional dip in capacitance is observed as the temperature is further reduced below ~10K, indicating opening of the mobility gap in the energy spectrum. **b.** The zero magnetic field conductance of 2D bulk as a function of temperature is measured in a Corbino disk, which vanishes exponentially with T. The conductance measured in Corbino disk at T = 300 mK are shown, respectively, for magnetic field applied in the plane in **c.** or perpendicular to the plane in **d.** In either case, there is no evidence for gap closing at increasing magnetic field; a continuous magnetic field sweep shows that 2D bulk is always completely insulating from 0 to 8T. **e.** The gap energy in the mobility gap is shown to increase with applied perpendicular magnetic field.

**Figure 4 a.** Plateau values measured under an inplane magnetic field for four different devices with in-plane magnetic field applied parallel (open circles) or perpendicular (open triangles) to the edge axis. **b.** shows the plateaus measured from the six-terminal device (shown in **a.** red open circles) at 20mK, at different applied in-plane magnetic field parallel to the edge. **c.** The same four samples were measured (T= 300 mK) in a magnetic field applied perpendicular to the 2D plane, with the three six-terminal devices showing incresing conductance, and the four-terminal device showing decreasing conductance. These observations are consistent with the notion that the edge transport characteristics change gradually from helical to chiral in a strong magnetic field, but still away from topological transition. **d.** shows that in the two-terminal device the edge resistance increases with a perpendicular magnetic field (0T, 2T, 4T, 6T, 8T) but remains smaller than $h/e^2$ in this magnetic field range.

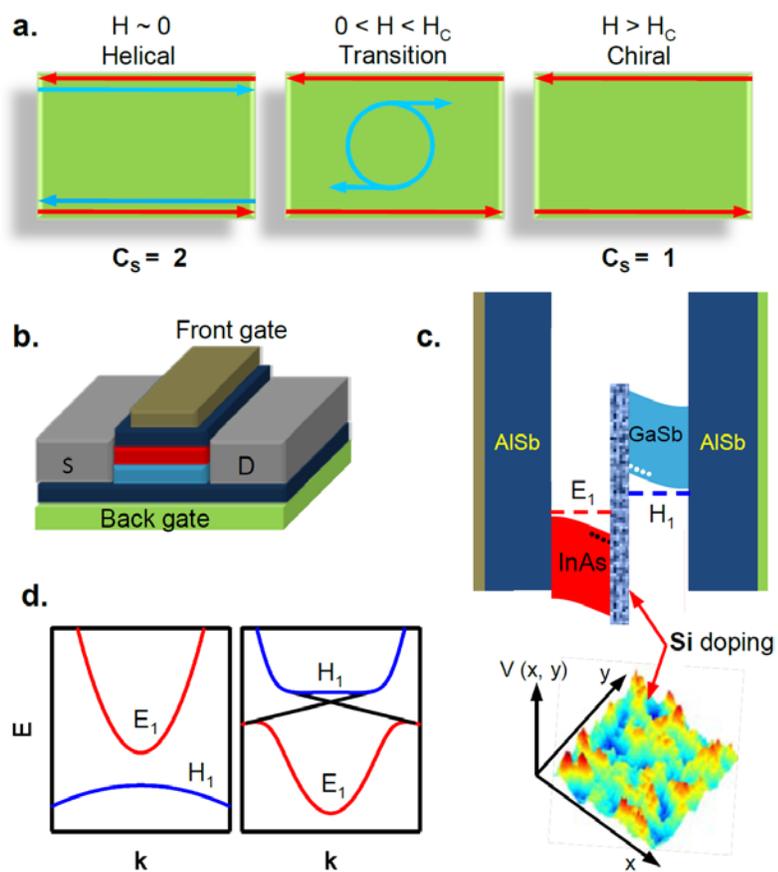

**Figure 1**

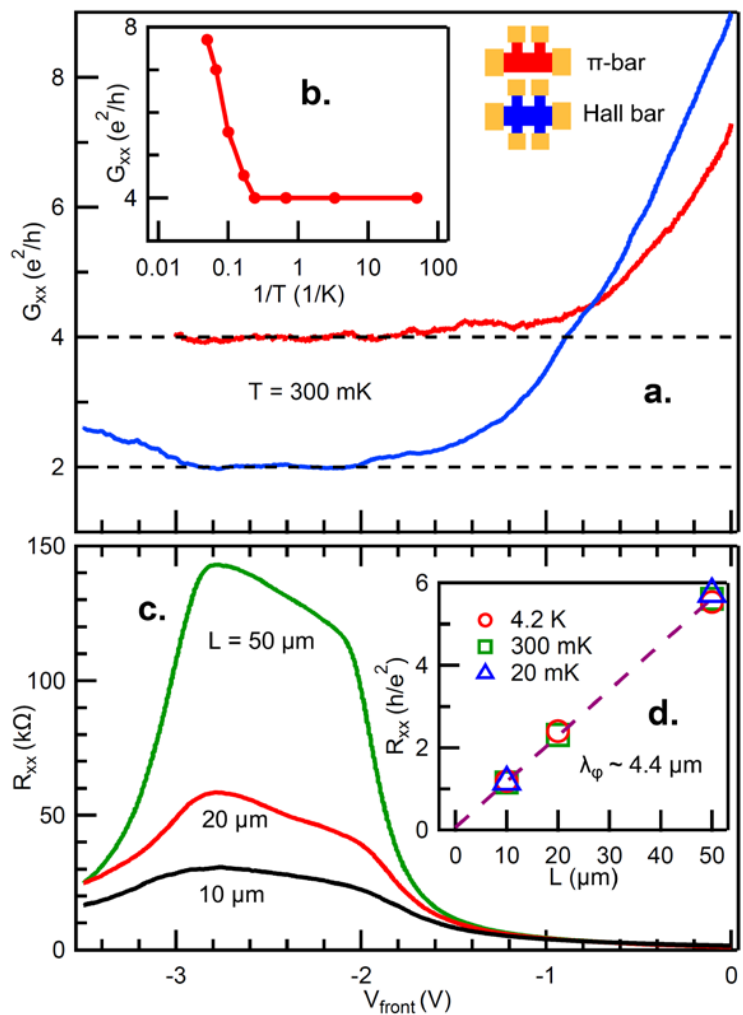

**Figure 2**

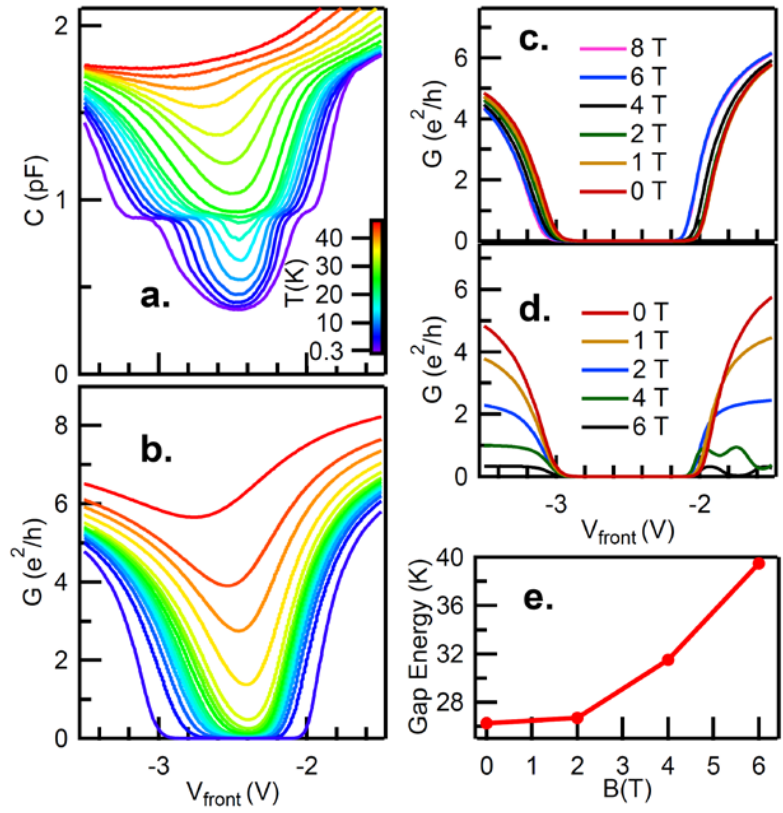

**Figure 3**

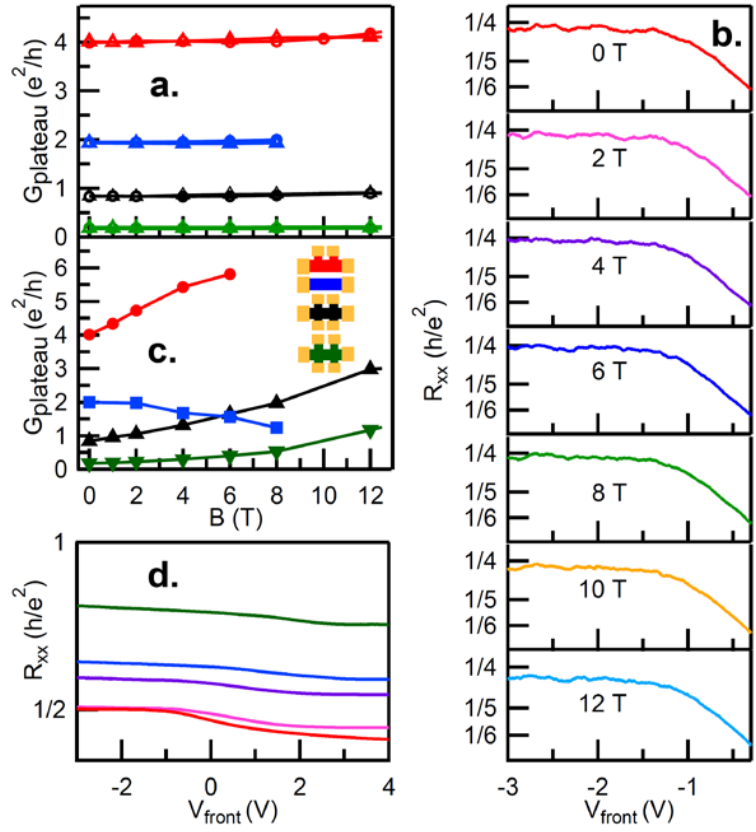

**Figure 4**